\documentclass[conference]{IEEEtran}
\usepackage[dvips]{graphicx}
\usepackage{multirow}
\usepackage{amsmath}
\usepackage{mathtools}
\usepackage{epstopdf}
\usepackage{authblk}
\usepackage{cite}
\usepackage{cleveref}
\usepackage{latexsym}
\usepackage[justification=centering]{caption}
\usepackage{amsthm}
\usepackage{balance}
\usepackage{relsize}
\usepackage[nodisplayskipstretch]{setspace}
\usepackage{color}
\usepackage{amssymb}
\usepackage{eucal}
\usepackage{lipsum}

\usepackage[nodisplayskipstretch]{setspace}
\usepackage{color}

\DeclareMathAlphabet\mathbfcal{OMS}{cmsy}{b}{n}

\begin{document}

\title{On the Performance of Non-Orthogonal Multiple Access Systems with Imperfect Successive Interference Cancellation}

\renewcommand\Authfont{\fontsize{12}{14.4}\selectfont}
\renewcommand\Affilfont{\fontsize{9}{10.8}\itshape}

\author[*]{Lina Bariah}
\author[*\dag]{Arafat Al-Dweik}
\author[* \ddag]{Sami Muhaidat}

\affil[*]{Department of Electrical and Computer Engineering, Khalifa University, Abu Dhabi, UAE. 
\protect\\  Emails: \{lina.bariah, dweik\}@kustar.ac.ae}
\affil[*\dag]{Western University, London, Canada. Email: dweik@fulbrightmail.org}
\affil[*\ddag]{University of Surrey, Guildford, U.K. Email: muhaidat@ieee.org}

\renewcommand\Authands{ and }

\maketitle
\begin{abstract}
\boldmath
Non-orthogonal multiple access (NOMA) technique has sparked a growing research interest due to its ability to enhance the overall spectral efficiency of wireless systems. In this paper, we investigate the pairwise error probability (PEP) performance of conventional NOMA systems, where an exact closed form expression for the PEP is derived for different users, to give some insight about the reliability of the far and near users. Through the derivation of PEP expressions, we demonstrate that the maximum achievable diversity order is proportional to the user's order. The obtained error probability expressions are used to formulate an optimization problem that minimizes the overall bit error rate (BER) under power and error rate threshold constrains. The derived analytical results, corroborated by Monte Carlo simulations, are presented to show the diversity order and error rate performance of each individual user.
\end{abstract}


\IEEEpeerreviewmaketitle

\section{Introduction}

Non-orthogonal multiple access (NOMA) is a promising technique for the upcoming fifth generation (5G) wireless communications, and it has attracted an increased research interests in recent years. Enhanced latency, spectral efficiency and connectivity are the main factors that stimulated the emergence of NOMA systems, in which multiple users are allowed to share the same time and frequency resources \cite{7842433}. The key point of NOMA systems is to permit a constrained level of interference from other users that allows the receiver to perform successive interference cancellation (SIC) for the other users' signals before detecting its own signal. NOMA systems rely on exploiting the power domain multiplexing to control interference and maintain user fairness, in a way that grants the far users higher power coefficients and assign low power coefficients to near users \cite{8010756}. Although NOMA technique enhances users' fairness, in comparison with the conventional systems such as orthogonal multiple access (OMA) systems, quality of service (QoS) of far users is relatively low, which is considered as a performance limiting factor in many scenarios due to error propagation. 

Extensive research efforts have been conducted to study the performance of NOMA systems from different perspectives and under different scenarios. In \cite{6868214}, the authors investigated the outage probability and the ergodic sum rates performance in downlink NOMA systems with randomly deployed users. The derived analytical results in \cite{6868214} show that the outage probability of NOMA systems highly depends on the targeted data rates and the allocated power for each user. Ding $\textit{et al.}$ \cite{7273963} studied the effect of user pairing on the outage probability performance and the sum rate for two scenarios, fixed power allocation and cognitive-radio inspired NOMA. As reported in \cite{7273963}, selecting users with distinctive channel gains can enhance the achieved sum rate. 

Dynamic power allocation for uplink and downlink NOMA systems is presented in \cite{7542118} with guaranteed QoS for different users. Unlike conventional techniques, such as fixed power allocation and cognitive-radio inspired NOMA, dynamic power allocation provides more flexibility by allowing tradeoffs between user fairness and overall system throughput. Performance analysis of NOMA systems is evaluated in \cite{7069272} from users' fairness standpoint. In particular, the authors investigate the outage probability and the sum rate of different power allocation scenarios, where instantaneous and average channel gains are considered. 

Although performance analysis of NOMA systems is well investigated in the literature \cite{7972963,7876764,7982626,7983401,8010439,7870605}, most of the reported work concentrates on evaluating the system's performance in terms of outage probability, individual sum rate and average sum rate. To the best of the authors knowledge, none of the reported work addressed the error rate performance analysis of NOMA systems. Emphasizing on this, studying the error rate performance of different users while considering imperfect SIC is crucial, to have some insightful results about the QoS of each individual user. Accurate bit error rate (BER) analysis of NOMA systems is intractable due to the SIC process, however, pairwise error probability (PEP) can be analyzed. It is worth noting that PEP gives a valuable indicator for the BER performance, since it is considered as an upper bound for the BER.

%
%

Based on the aforementioned discussion, the main contributions of this paper are summarized as follows:
\begin{itemize}
\item In this work, the PEP performance analysis of conventional NOMA systems with imperfect SIC is considered, where an exact closed form PEP expression is derived for each user individually. The derived PEP expressions are verified by Monte Carlo simulations.
\item Building on the obtained PEP formulae, asymptotic PEP is derived to analyze the achieved effective diversity gain, which represents the performance of the system at high SNR regime.
\item Using the derived asymptotic expression of the PEP, an optimization problem is formulated and solved to obtain the optimum power allocation coefficients that minimize the BER, under power and users' individual error rate constrains.
\end{itemize}
The rest of the paper is organized as follows. Adopted system and channel models are presented in Sec. \ref{sec:models} followed by exact and asymptotic PEP analysis for each individual user in Sec. \ref{sec:pep}. Power allocation coefficients optimization is addressed in Sec. \ref{sec:pwr}. Numerical and simulation results are presented in Sec. \ref{sec:results2} and the paper is concluded in Sec. \ref{sec:con}.

\textit{Notation:} $(\cdot )^{\ast}$ and $\left |. \right |$ denote the complex conjugate operation and the absolute value, respectively. $\textup{Re}\left\{.\right\}$ represents the real part of a complex number. $\hat{x}$ represents a detected symbol and $\Delta$ denotes $(x - \hat{x})$. 

\section{System and Channel Models}\label{sec:models}

\begin{figure}[ht]
\centering
\includegraphics[width=0.96\linewidth]{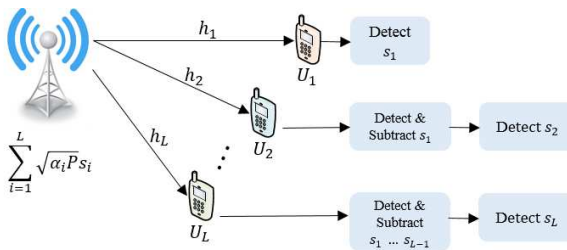}
\caption{Typical NOMA system with $L$ users.}
\label{fig:NOMA}
\end{figure}

Recalling that the basic idea behind NOMA systems is to utilize the broadcast nature of the wireless channels to allow multiple users to share the same time, frequency and code domains while assigning different power levels for different users, to permit a specific level of interference from the other users. In this work, downlink transmission NOMA system with $L$ users is considered, where each user is equipped with single antenna, as depicted in Fig. \ref{fig:NOMA}. Users are classified based on their distance from the base station (BS), where the first user is the farthest user from the BS, consequently, it has the weakest channel. On the other hand, the $L$th user is the nearest with the strongest channel. The channels between the BS and the $L$ users are modeled as independent and identically distributed (i.i.d) Rayleigh flat fading channels. It is worth mentioning that near users are assigned lower power coefficients than far users. Given the total transmitted signal power is $P$, the transmitted signal from the BS is given by,
\begin{equation}
s = \sum_{l=1}^{L}\sqrt{\alpha_{l}P} \;x_{l}
\end{equation}
\noindent where $x_{l}$ is the transmitted signal of the $l$th user and $\alpha_{l}$ is the power allocation coefficient, where $\sum_{l=1}^{L}\alpha_{l} =1$. The received signal at the $l$th user is,
\begin{equation}
r_{l} = h_{l} \;s + n_{l}
\end{equation} 
\noindent where $h_{l} \sim \CMcal{CN}(0,2\sigma_{h}^{2})$ is the channel frequency response and $n_{l}$ is the additive white Gaussian noise (AWGN) with zero mean and variance $\sigma_{n}^{2}$. Power allocation coefficients are sorted in descending order, $\alpha_{1} > \alpha_{2} > ... > \alpha_{L}$, given that $|h_{1}|^{2}<|h_{2}|^{2}<...<|h_{L}|^{2}$. The first user decodes only its signal $x_{1}$, while treating the signals of all other users as interference. The rest of the users should employ SIC to be able to detect their signals. For the $l$th user, it should perform SIC for the higher power users, i.e., $U_{1}, \cdots U_{l-1}$, and treat the rest of users signals as interference, i.e., $U_{l+1}, \cdots U_{L}$. 

\section{Pairwise Error Probability Analysis for NOMA Systems} \label{sec:pep}

\subsection{PEP Analysis for First User}

Without loss of generality, we consider the first user as the farthest user, therefore, $\left | h_{1} \right |^{2} < \left | h_{2} \right |^{2} < \cdots <\left | h_{L} \right |^{2}$. The received signal at the first user can be represented as follows,
\begin{equation}
\label{eq:r1}
r_{1} = h_{1}\left ( \sqrt{\alpha_{1}P}\; x_{1} + \sum_{l=2}^{L}\sqrt{\alpha_{l}P}\; x_{l}\right ) + n_{1}
\end{equation}
\noindent where $\sum_{l=2}^{L}\sqrt{\alpha_{l}P}\; x_{l}$ represents the interference term from the other users. PEP is defined as the probability of detecting the symbol $\hat{x}$ while symbol $x$ was transmitted \cite{1603372}, which can be evaluated for the first user as follows,
 \begin{equation}
\label{eq:pep1}
\begin{split}
&\textup{PEP}\left ( x_{1},\hat{x}_{1} \right )=\\ &\textup{Pr}\left (  \left | r_{1} -\sqrt{\alpha_{1}P}\;h_{1} \hat{x}_{1}\right |^{2} \leq  \left | r_{1} -\sqrt{\alpha_{1}P}\;h_{1} x_{1}\right |^{2} \right ), \hat{x}_{1} \neq x_{1}.
\end{split}
\end{equation}
Using the cumulative distribution function (CDF) of a normal distribution, the conditional PEP for the first user can be represented as given in (\ref{eq:pep3}). 
\begin{figure*}[ht]
\begin{equation}
\label{eq:pep3}
\textup{PEP}\left ( x_{1},\hat{x}_{1} \mid |h_1|\right )=\textup{Q}\left ( \frac{\sqrt{\alpha_{1}P}\;\left | h_{1} \right |\left | \Delta_{1}   \right |^{2} + 2\left | h_{1} \right |\textup{Re}\left \{ \Delta_{1}\sum_{l=2}^{L}\sqrt{\alpha_{l}P}\; x_{l}^{\ast}   \right \}}{\sqrt{2}\left | \Delta_{1} \right | \sigma_{n}} \right ). 
\end{equation}
\hrulefill
\vspace*{4pt}
\end{figure*}
\noindent In (\ref{eq:pep3}),
\begin{equation}
\textup{Q}(x)=\frac{1}{\sqrt{2 \pi}}\int_{x}^{\infty} \textup{exp}\left ( -\frac{u^{2}}{2} \right )du
\end{equation} 
\noindent is the Gaussian Q-function \cite{porakis} and $\Delta_{1} = (x_{1} - \hat{x}_{1})$. It is worth noting that the derived PEP expressions are conditioned on particular interference values, which depend on the transmitted and detected symbols for each user. 

To get the unconditional PEP, we average over the probability density function (PDF) of $|h|$. By noting that user 1 has always the weakest channel, and channel gains for the rest of users are ordered in ascending order, i.e. $|h_{1}| = \textup{min}(|h_{1}|, \cdots,|h_{L}|)$ and  $|h_{L}| = \textup{max}(|h_{1}|, \cdots,|h_{L}|)$, ordered statistics should be considered when evaluating the PDF of $|h_{1}|$. Therefore, the PDF of the $l$th user is given by \cite{7510798},
\begin{equation}
\label{eq:pdf}
f_{(l)}(x)=\frac{L!}{(l-1)!(L-l)!}f_{X}(x)F_{X}(x)^{l-1}\left ( 1-F_{X}(x) \right )^{L-l}.
\end{equation}
Considering that $|h|$ is Rayleigh distributed, its PDF and CDF are $f_{X}(x) = \frac{x}{\sigma^{2}}\textup{exp}\left ( -\frac{x^{2}}{2\sigma^{2}} \right )$ and $F_{X}(x) =1- \textup{exp}\left ( -\frac{x^{2}}{2\sigma^{2}} \right )$, respectively \cite{ALO}. Therefore, using (\ref{eq:pdf}), the PDF of $\left | h_{1} \right | \triangleq \omega_{1}$ is given by,
\begin{equation}
\label{eq:pdfh1}
f_{\Omega}(\omega_{1}) =\frac{2\omega_{1}}{\sigma_{h}^{2}}\;\textup{exp}\left ( -\frac{\omega_{1}^{2}}{2\sigma_{h}^{2}} \right ) 
\end{equation}
\noindent where $\sigma_{h}^{2} = \mathcal{E}\left [ \left | h_{l} \right |^{2} \right ]$, $l=1,2,\cdots,L$.

\noindent Hence, the PEP averaged over the PDF of $\omega_{1}$ is
\begin{equation}
\label{eq:pep5}
\textup{PEP}\left (x_{1},\hat{x}_{1} \right )=\int_{0}^{\infty}\frac{\omega_{1}}{\sigma_{h}^{2}}\;\textup{exp}\left ( -\frac{\omega_{1}^{2}}{2\sigma_{h}^{2}} \right )\textup{erfc} \left ( \frac{\Gamma \; \omega_{1}}{\sqrt{2}\;\zeta} \right )d\omega_{1}
\end{equation}
\noindent where
\begin{equation}
\label{eq:gamma}
\Gamma = \sqrt{\alpha_{1}P}\;\left | \Delta_{1}   \right |^{2} + 2\textup{Re}\left \{ \Delta_{1}\sum_{l=2}^{L}\sqrt{\alpha_{l}P}\; x_{l}^{\ast}   \right \}
\end{equation}
\noindent and
\begin{equation}
\label{eq:zeta}
\zeta = \sqrt{2}\left | \Delta_{1} \right | \sigma_{n}.
\end{equation}
In (\ref{eq:pep5}), we use the identity, $\textup{Q}(x) = \frac{1}{2}\textup{erfc}(\frac{x}{\sqrt{2}})$, where $\textup{erfc}(x)$ is the complementary error function. Solving the integral in (\ref{eq:pep5}) gives \cite{int2},
\begin{equation}
\label{eq:pep6}
\textup{PEP}\left ( x_{1},\hat{x}_{1} \right )= \frac{1}{2}\left (1-\frac{\Gamma \sigma_{h}}{\sqrt{2 \zeta^{2}+\Gamma^{2}\sigma_{h}^{2}}}  \right ).
\end{equation}
\noindent which can be averaged over all the possible values of $x_{l}, l=2,\cdots,L$, to consider all interference scenarios.

\subsection{PEP Analysis for the $l$\textup{th} User}

For the $l$th user, it first decodes the signals with higher power, i.e., $U_{1}, \cdots, U_{l-1}$, to perform SIC before detecting its own signal. The output of the $l$th SIC receiver can be represented as,
\begin{equation}
\label{eq:rl}
\tilde{r}_{l} = \sqrt{\alpha_{l}P}\; h_{l}x_{l}+ \sum_{n=l+1}^{L}\sqrt{\alpha_{n}P}h_{l}x_{n}+\sum_{k=1}^{l-1}\sqrt{\alpha_{k}P}\;h_{l}\Delta_{k} + n_{l}
\end{equation}
\noindent where $\Delta_{k} = (x_{k} - \hat{x}_{k})$. The PEP of the $l$th user can be evaluated as shown in (\ref{eq:pep1}), which after simplification can be represented as shown in (\ref{eq:pep8}). We would like to highlight that for the $L$th user, the term $\textup{Re}\left \{ \Delta_{l}\sum_{n=l+1}^{L}\sqrt{\alpha_{n}P}\;x_{n}^{\ast} \right \}$ equals to zero. Hence, the PEP of the $L$th user is given in (\ref{eq:pep81}).
\begin{figure*}[ht]
\small
\begin{equation}
\label{eq:pep8}
\textup{PEP}\left ( x_{l},\hat{x}_{l}| \left | h_{l} \right | \right )= \textup{Pr}\left ( 2\sqrt{\alpha_{l}P}\; \textup{Re}\left \{ h_{l} \Delta_{l}n_{l}^{\ast} \right \} \leq -h_{l}^{2}\left ( \alpha_{l}P\left | \Delta_{l} \right |^{2}+2\sqrt{\alpha_{l}P}\; \left [ \textup{Re}\left \{ \Delta_{l}\sum_{n=l+1}^{L}\sqrt{\alpha_{n}P}\;x_{n}^{\ast} \right \}+\textup{Re}\left \{ \Delta_{l}\sum_{k=1}^{l-1}\sqrt{\alpha_{k}P}\;\Delta_{k}^{\ast} \right \} \right ]  \right ) \right ). 
\end{equation}
\normalsize
\hrulefill
\vspace*{4pt}
\end{figure*}
\begin{figure*}[ht]
\small
\begin{equation}
\label{eq:pep81}
\textup{PEP}\left ( x_{L},\hat{x}_{L}| \left | h_{L} \right | \right )= \textup{Pr}\left ( 2\sqrt{\alpha_{L}P}\; \textup{Re}\left \{ h_{L} \Delta_{L}n_{L}^{\ast} \right \} \leq -h_{L}^{2}\left ( \alpha_{L}P\left | \Delta_{L} \right |^{2}+2\sqrt{\alpha_{L}P}\; \textup{Re}\left \{ \Delta_{L}\sum_{k=1}^{L-1}\sqrt{\alpha_{k}P}\;\Delta_{k}^{\ast} \right \} \right ) \right ). 
\end{equation}
\normalsize
\hrulefill
\vspace*{4pt}
\end{figure*}
Therefore, using the CDF of a normal Gaussian random variable, the conditional PEP of the $l$th user can be evaluated as the following,
\begin{equation}
\label{eq:pep9}
\textup{PEP}\left ( x_{l},\hat{x}_{l}| \left | h_{l} \right | \right )= \textup{Q}\left (  \frac{\left |h_{l}\right | \beta_{l}}{\upsilon } \right )
\end{equation}
\noindent where
\begin{equation}
\label{eq:beta}
\begin{split}
\beta_{l} = \sqrt{\alpha_{l}P}\left | \Delta_{l} \right |^{2}&+2 \Bigg [ \textup{Re}\left \{ \Delta_{l}\sum_{n=l+1}^{L}\sqrt{\alpha_{n}P}\;x_{n}^{\ast} \right \} \\ &+\textup{Re}\left \{ \Delta_{l}\sum_{q=1}^{l-1}\sqrt{\alpha_{q}P}\;\Delta_{q}^{\ast} \right \} \Bigg ]
\end{split}
\end{equation}
\noindent and
\begin{equation}
\label{eq:v}
\upsilon =\sqrt{2}\sigma_{n}\left | \Delta_{l} \right |.
\end{equation}

\noindent To evaluate the unconditional PEP, we average over the PDF of $ \left | h_{l} \right | \triangleq \omega_{l}$. Using the PDF of the ordered statistics provided in (\ref{eq:pdf}) and considering that $\left | h \right |$ is Rayleigh distributed, the PDF of $\left | h_{l} \right |$ is,
\begin{equation}
\label{eq:pdfL}
\begin{split}
f_{\Omega}(\omega_{l})=&\frac{L!}{(l-1)!(L-l)!}\; \frac{\omega_{l}}{\sigma_{h}^{2}}\; \textup{exp}\left ( -\frac{\omega_{l}^{2}}{2\sigma_{h}^{2}} \right )\\
&\left ( 1-\textup{exp}\left ( -\frac{\omega_{l}^{2}}{2\sigma_{h}^{2}} \right ) \right )^{l-1}\left ( \textup{exp}\left ( -\frac{\omega_{l}^{2}}{2\sigma_{h}^{2}} \right ) \right )^{L-l}.
\end{split}
\end{equation}
To calculate the unconditional PEP, we use binomial expansion $\left ( a+x \right )^{n}=\sum_{k=0}^{n}\binom{n}{k}x^{k}a^{n-k}$ \cite[Eq. 1.111]{IntTable} to represent the term $\left ( 1-\textup{exp}\left ( -\frac{\omega_{l}^{2}}{2\sigma_{h}^{2}} \right ) \right )^{l-1}$. Accordingly, the PEP can be evaluated using the following integral,
\begin{equation}
\label{eq:pep11}
\begin{split}
\textup{PEP}&\left ( x_{l},\hat{x}_{l} \right ) =\frac{L!}{\sigma_{h}^{2}(l-1)!(L-l)!}\sum_{j=0}^{l-1}\binom{l-1}{j}\left ( -1 \right )^{2(l-1)-j}\\
& \times \int_{0}^{\infty}\omega_{l} \; \textup{exp}\left ( -\frac{\left [ L-l+j-1 \right ]\omega_{l}^{2}}{2\sigma_{h}^{2}} \right )\textup{Q}\left ( \frac{\beta_{l} \omega_{l}}{\upsilon } \right )d\omega_{l}.
\end{split}
\end{equation} 
Solving the integral in (\ref{eq:pep11}) gives the closed form expression for the PEP for the $l$th user, as shown in (\ref{eq:pep12}).   
\begin{figure*}[ht]
\begin{equation}
\label{eq:pep12}
\textup{PEP}\left ( x_{l},\hat{x}_{l} \right ) =\frac{L!}{\sigma_{h}^{2}(l-1)!(L-l)!}\sum_{j=0}^{l-1}\binom{l-1}{j}\frac{\left ( -1 \right )^{2(l-1)-j}}{\left [ L-l+j+1 \right ]}\left (1-\frac{\beta_{l} \sigma_{h}}{\sqrt{\beta_{l}^{2}\sigma_{h}^{2}+\left [ L-l+j+1 \right ]\upsilon^{2}}}  \right ). 
\end{equation}
\hrulefill
\vspace*{4pt}
\end{figure*}

\subsection{Asymptotic Analysis}
PEP represents an upper bound for the BER, and it gives a useful insight on the error rate performance when the closed form expression of the BER can not be found. PEP is used also to study the achieved diversity, where the diversity gain is defined as the magnitude of the slope of the PEP when the signal-to-noise ratio (SNR) value goes to infinity \cite{1603372},
\begin{equation}
\label{eq:Asymp}
d_{s} = \lim_{\bar{\gamma}\rightarrow \infty } -\frac{\textup{log}\;\textup{PEP}\left ( x_{l},\hat{x}_{l} \right )}{\textup{log}\;\bar{\gamma}}
\end{equation} 
\noindent where $\bar{\gamma}=\CMcal{E}\left \{  \gamma \right \}$ is the average transmit SNR. Capitalizing on the PEP presented in (\ref{eq:pep12}), in this section we derive the asymptotic expression for the PEP of the $l$th user, which will be used to evaluate the asymptotic diversity order. In this work we will concentrate on the effective diversity gain,
\begin{equation}
\label{eq:Asymp2}
d_{e} =  -\frac{\textup{log}\;\textup{PEP}\left ( x_{l},\hat{x}_{l} \right )}{\textup{log}\; \bar{\gamma}}.
\end{equation}
\noindent As it is noticed, when $\bar{\gamma} \rightarrow \infty$, the effective diversity order converges to the asymptotic diversity gain. The conditional PEP presented in Eqn. (\ref{eq:pep9}) can be bounded by the following,
\begin{equation}
\label{eq:pepAs}
\textup{PEP}\left ( x_{l},\hat{x}_{l}| \left | h_{l} \right | \right ) \leq \textup{exp}\left (  -\frac{\gamma \beta_{l}^{2}}{4\left | \Delta_{l} \right |^{2}} \right )
\end{equation} 
\noindent where $\beta_{l}$ is given in (\ref{eq:beta}) and $\gamma = \left | h_{l} \right |^{2}/\sigma_{n}^{2}$ is the instantaneous SNR, which is modeled as exponential random variable with PDF,
\begin{equation}
\label{eq:pdfExp}
f(\gamma)=\frac{1}{\bar{\gamma}}\textup{exp}\left ( -\frac{\gamma}{\bar{\gamma}} \right ).
\end{equation}  
\noindent Using (\ref{eq:pdfExp}) and the ordered statistics PDF provided in (\ref{eq:pdf}) and after some manipulations, the ordered PDF of the instantaneous SNR at the $l$th user is given by,
\begin{equation}
\label{eq:pdfSNR}
f_{l}(\gamma)=A_{l}\sum_{j=0}^{l-1}\binom{l-1}{j}\left ( -1 \right )^{j}\frac{1}{\bar{\gamma}}\; \left [\textup{exp}\left ( -\frac{\gamma}{\bar{\gamma}} \right )  \right ]^{j+L-l+1}
\end{equation}  
\noindent where $A_{l}=\frac{L!}{(l-1)!(L-l)!}$. 

Therefore, the asymptotic unconditional PEP can be evaluated as,
\begin{equation}
\label{eq:pepAs2}
\begin{split}
\textup{PEP}\left ( x_{l},\hat{x}_{l}\right ) & \leq A_{l}\sum_{j=0}^{l-1}\binom{l-1}{j}\left ( -1 \right )^{j}\frac{1}{\bar{\gamma}}\times \\
& \int_{0}^{\infty}\left [\textup{exp}\left ( -\frac{\gamma}{\bar{\gamma}} \right )  \right ]^{j+L-l+1}\textup{exp}\left (  -\frac{\gamma \beta_{l}^{2}}{4\left | \Delta_{l} \right |^{2}} \right )d\gamma.
\end{split}
\end{equation}
Given that the diversity order is evaluated at high SNR values, the first exponential in (\ref{eq:pepAs2}) can be approximated as $\textup{exp}\left ( -\frac{\gamma}{\bar{\gamma}} \right )\approx (1-\frac{\gamma}{\bar{\gamma}})$. 

Hence, 
\begin{equation}
\label{eq:pepAs3}
\begin{split}
\textup{PEP}\left ( x_{l},\hat{x}_{l}\right ) & \leq A_{l}\sum_{j=0}^{l-1}\binom{l-1}{j}\left ( -1 \right )^{j}\frac{1}{\bar{\gamma}}\times \\
& \int_{0}^{\infty}\left (1 -\frac{\gamma}{\bar{\gamma}} \right ) ^{j+L-l+1}\textup{exp}\left (  -\frac{\gamma \beta_{l}^{2}}{4\left | \Delta_{l} \right |^{2}} \right )d\gamma.
\end{split}
\end{equation}
Solving the integral in (\ref{eq:pepAs3}) and after some simplifications, the bounded PEP can be expressed as follows,
\begin{equation}
\label{eq:pepAs4}
\begin{split}
\textup{PEP}\left ( x_{l},\hat{x}_{l}\right ) \leq \frac{A_{l}}{\bar{\gamma}}\sum_{j=0}^{l-1}\sum_{k=0}^{z}\binom{l-1}{j}& \binom{z}{k}(-1)^{j+z+k}(\bar{\gamma})^{-z+k} \\ 
&\Gamma(z-k+1) \left (\frac{4\left | \Delta_{l} \right |^{2}}{\beta_{l}^{2}}  \right ) 
\end{split}
\end{equation}
\noindent where $z = j+L-l+1$. At high SNR values and considering the dominant components from the summations in (\ref{eq:pepAs4}), it is observed that the bounded PEP is proportional to the effective diversity order, 
\begin{equation}
\textup{PEP}\left ( x_{l},\hat{x}_{l}\right ) \propto \bar{\gamma}^{-z+k-1}. 
\end{equation}
The effective diversity order is evaluated from (\ref{eq:pepAs4}) using numerical methods and results are provided in Sec. \ref{sec:results2}.

\section{Power Allocation Coefficients Optimization}
\label{sec:pwr}

It has been demonstrated in literature and using numerical and analytical results, that power allocation coefficients play an essential rule in determining the overall performance of the NOMA systems. Proper power allocation among different users can enhance the overall performance remarkably. In this section, we will form an optimization problem that aims to find the optimum power allocation coefficients that minimizes the average BER. It is worth mentioning that PEP is used to calculate a union bound on the BER, as follows \cite{porakis},
\begin{equation}
\label{eq:BER}
P_{e} \leq \sum_{m=1}^{M}P_{m}\sum_{\substack{\tilde{m}=1\\ x \neq\hat{x}}}^{M}q(x_{(m)}\rightarrow \hat{x}_{(\hat{m})}) \textup{PEP}(x_{(m)},\hat{x}_{(\hat{m})})
\end{equation}
\noindent where $P_{m}$ is the probability that $x_{(m)}$ is transmitted and $q(x_{(m)}\rightarrow \hat{x}_{(\hat{m})})$ is the number of bit errors between $x_{(m)}$ and $\hat{x}_{(\hat{m})}$. Therefore, our aim is to find the optimum power allocation coefficients that minimize the following objective function,
\begin{equation}
\label{eq:obj}
\Psi =  \sum_{m=1}^{M}P_{m}\sum_{\substack{\tilde{m}=1\\ x \neq\hat{x}}}^{M}q(x_{(m)}\rightarrow \hat{x}_{(\hat{m})}) \textup{PEP}(x_{(m)},\hat{x}_{(\hat{m})})
\end{equation} 
\noindent while satisfying a specific error rate performance threshold for all users to maintain user fairness. Additionally, for normalized average power, the some of the power allocation coefficients should equals to 1. Hence, the optimization problem can be represented as,
\begin{equation}
\label{eq:opt}
\begin{matrix}
\textup{Minimize} & \Psi \\ 
s.t. & \left\{\begin{matrix}
\sum_{j=1}^{L}\alpha_{j}=1, \\ \textup{PEP}(x_{l},\hat{x_{l}}) \leq P_{th}.
\end{matrix}\right.
\end{matrix}
\end{equation} 
The above optimization problem is solved using numerical methods since closed form expressions for the optimum coefficients are hard to derive.

\section{Numerical and Simulation Results}
\label{sec:results2}

In this section, numerical and simulation results are conducted to evaluate the performance of the proposed scheme and to validate the derived analytical results. A conventional NOMA system is adopted where a single BS and three users are considered with power allocation coefficients $\alpha_{1}$, $\alpha_{2}$ and $\alpha_{3}$, for the first, second and third user, respectively. Without loss of generality, we consider the first user as the farthest user, $\alpha_{1}>\alpha_{2}>\alpha_{3}$. All users are equipped with single antenna and the link between each user and the BS is considered as Rayleigh flat fading channel. Transmitted signals are chosen randomly from quadrature phase shift keying (QPSK) constellation with average power $P = 1$. It is worth mentioning that in the presented results, the transmitted signals of different users are fixed and imperfect SIC is considered.

\begin{figure}[ht]
\centering
\includegraphics[width=1\linewidth]{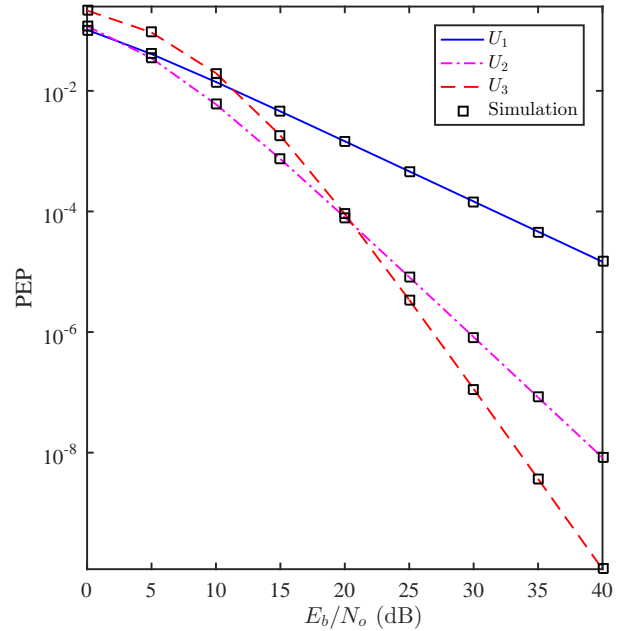}
\caption{Analytical and simulated PEP for the 3 users with imperfect SIC.}
\label{fig:PEP1}
\end{figure}

Fig. \ref{fig:PEP1} presents the PEP for the three users while considering imperfect SIC scenarios. Power allocation coefficients are $\alpha_{1} = 0.7$, $\alpha_{2} = 0.2$ and $\alpha_{3} = 0.1$. This power allocations coefficients values are chosen based on the evaluated performance of the system, where it is noted that these values give good performance in comparison with other values. The derived analysis are corroborated with simulation results, where it is shown that the derived analysis and simulation results match perfectly for the three users over the entire SNR range. As expected, the PEP gives an indication about the performance of the three users in NOMA systems in low and high SNR values, where at high SNR value, the near users show strong performance while the far user has relatively weak performance.

\begin{figure}[ht]
\centering
\includegraphics[width=1\linewidth]{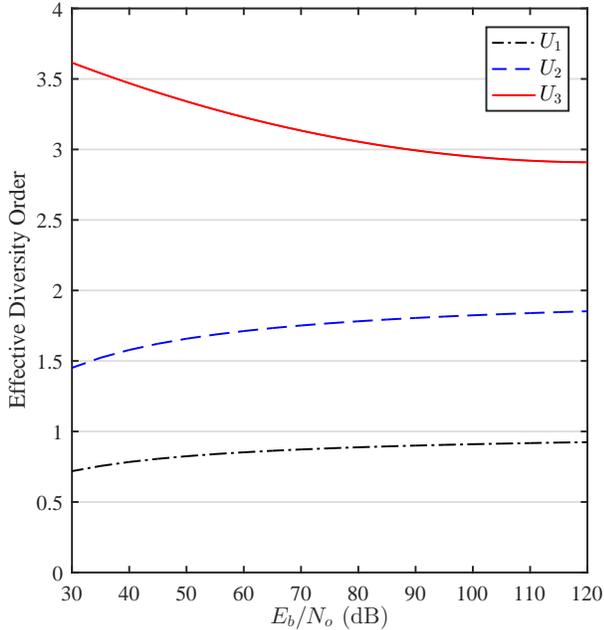}
\caption{Effective diversity order for the three users, $\alpha_{1} = 0.7$, $\alpha_{2} = 0.2$ and $\alpha_{3} = 0.1$.}
\label{fig:Div}
\end{figure}

The effective diversity order of different users is shown in Fig. \ref{fig:Div}. From the figure, it is observed that at high SNR values, the diversity order of the $l$th user converges to $l$. Which is expected since the asymptotic diversity gain is achieved when the PEP of NOMA systems behaves as $\textup{PEP}(x,\hat{x})\propto \bar{\gamma}^{-d_{e}}$ \cite{1603372}. It is noted here that diversity gain in NOMA systems is realized due to the ordered channel gains, which in reality represents how far each user from the BS. 

\begin{figure}[ht]
\centering
\includegraphics[width=1\linewidth]{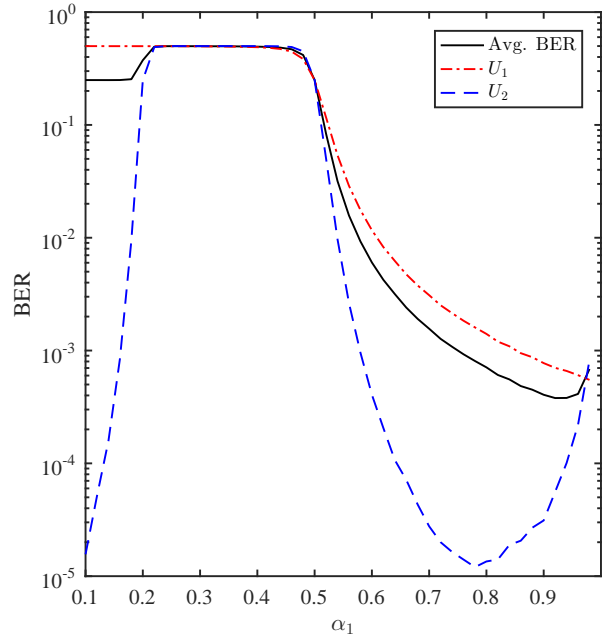}
\caption{Average BER over different power allocations, SNR = 30 dB, $\alpha_{2} = 1 - \alpha_{1}$, $P_{th} = 10^{-3}$.}
\label{fig:Opt}
\end{figure}

Fig. \ref{fig:Opt} shows the average and individual error rate performance of NOMA system with two users scenario over different combinations of power allocation coefficients, where SNR = 30 dB. From the figure, it is noticed that the second user can achieve the threshold error rate at very low and very high values of $\alpha_{1}$. However, at very low values of $\alpha_{1}$, the first user has a very poor performance, and this is justified by the increased interference from the second user. Although the second user achieves the best performance when $\alpha_{1} = 0.7781$, at this value of $\alpha_{1}$ the first user exceeds the threshold value, where $P_{th} = 10^{-3}$, hence, user fairness is violated in this scenario. To achieve users' fairness, where both users have error rate performance less than the threshold value while the average BER is kept to the minimum, $\alpha_{1}$ should take values from 0.852 to 0.99. Choosing the optimum power allocation coefficients is a tradeoff problem that is determined based on the targeted average BER and the individual BER of each user.

\section{Conclusion}
\label{sec:con}

In this paper, we investigated the performance of NOMA systems from error rate standpoint. An exact closed form expression for the PEP is derived, which represents a tight upper bound for the BER, therefore, it can give useful indication about the BER performance of each user in NOMA systems. 
Using the obtained PEP, asymptotic expression is derived, which is then used to evaluate the achieved effective diversity order. Capitalizing on the importance of the allocated power coefficients, constrained optimization problem is introduced to evaluate the optimum coefficients that reduce the overall error rate. Derived expressions, verified by Monte Carlo simulation results, gave an insightful results about the users' reliability and error rate performance.  

\section{Acknowledgment}
This work was supported by ICT fund grant No. 11/15/TRA-ICTFund/KU.

\bibliographystyle{IEEEtran}
\bibliography{references}

\end{document}